\documentclass[12pt]{article}
\usepackage{graphicx}
\def \babar{B{\sc a}B{\sc ar}}
\def\beq{\begin{equation}}
\def\eeq{\end{equation}}
\def\bea{\begin{eqnarray}}
\def\eea{\end{eqnarray}}
\def\bwt{\begin{widetext}}
\def\ewt{\end{widetext}}
\def\nn{\nonumber}
\def\roughly#1{\mathrel{\raise.3ex\hbox
{$#1$\kern-.75em\lower1ex\hbox{$\sim$}}}}

\def\ks{K_S}
\def\kbar{{\bar K}^0}
\def\bd{B^0}
\def\bdbar{{\bar B}^0}

\def\btos{{\bar b} \to {\bar s}}
\def\order{\lower 1.8ex \hbox{\LARGE\~{}}}

\def\btokpipi{B \to K \pi \pi}
\def\btokkk{B \to KK{\bar K}}

\begin{document}

\begin{flushright}
UdeM-GPP-TH-13-220 \\
\end{flushright}

\begin{center}
\bigskip
{\Large \bf \boldmath Extraction of the CP-violating phase $\gamma$
  using $\btokpipi$ and $\btokkk$ decays} \\
\bigskip
\bigskip
{\large
Bhubanjyoti Bhattacharya $^{a,}$\footnote{bhujyo@lps.umontreal.ca},
Maxime Imbeault $^{b,}$\footnote{mimbeault@cegep-st-laurent.qc.ca} \\
and David London $^{a,}$\footnote{london@lps.umontreal.ca}
}
\end{center}

\begin{flushleft}
~~~~~~~~~~~$a$: {\it Physique des Particules, Universit\'e
de Montr\'eal,}\\
~~~~~~~~~~~~~~~{\it C.P. 6128, succ. centre-ville, Montr\'eal, QC,
Canada H3C 3J7}\\
~~~~~~~~~~~$b$: {\it D\'epartement de physique, C\'egep de Saint-Laurent,}\\
~~~~~~~~~~~~~~~{\it 625, avenue Sainte-Croix, Montr\'eal, QC, Canada H4L 3X7 }
\end{flushleft}

\begin{center}
\bigskip (\today)
\vskip0.5cm {\Large Abstract\\} \vskip3truemm
\parbox[t]{\textwidth}{Using the \babar\ measurements of the Dalitz
  plots for $\bd \to K^+\pi^0\pi^-$, $\bd \to K^0\pi^+\pi^-$, $B^+ \to
  K^+\pi^+\pi^-$, $\bd \to K^+ K^0 K^-$, and $\bd \to K^0 K^0 \kbar$
  decays, we demonstrate that it is possible to cleanly extract the
  weak phase $\gamma$. We find four possible solutions. Three of these
  -- $32^\circ$, $259^\circ$, and $315^\circ$ -- are in disagreement
  with the SM, while one -- $77^\circ$ -- is consistent with the SM.
  An advantage of this Dalitz-plot method is that one can obtain many
  independent measurements of $\gamma$, thereby reducing its
  statistical error. An accurate determination of the errors, however,
  requires detailed knowledge of the data.}

\end{center}

\thispagestyle{empty}
\newpage
\setcounter{page}{1}
\baselineskip=14pt

One of the main aims of $B$ physics is to test the standard model (SM)
explanation of CP violation, which is that it is due to a complex
phase in the Cabibbo-Kobayashi-Maskawa (CKM) quark mixing matrix. To
this end, one measures the three angles of the unitarity triangle
\cite{pdg}, $\alpha$, $\beta$ and $\gamma$, in many different ways,
and looks for discrepancies.

The conventional wisdom has been that one can cleanly extract CKM
phase information only from two-body $B$ decays. However, it was
recently shown that, contrary to this point of view, such information
can also be obtained from charmless three-body $B$ decays
\cite{3body1,3body2}. Based on this result, a method was proposed for
extracting the weak phase $\gamma$ from $\btokpipi$ and $\btokkk$
decays \cite{3body3}. Specifically, $\gamma$ is obtained by combining
information from the Dalitz plots for $\bd \to K^+\pi^0\pi^-$, $\bd
\to K^0\pi^+\pi^-$, $B^+ \to K^+\pi^+\pi^-$, $\bd \to K^+ K^0 K^-$,
and $\bd \to K^0 K^0 \kbar$.

In this paper, we apply this method to experimental data. We use the
measurements of the Dalitz plots of the five $\btokpipi$ and $\btokkk$
decays by the \babar\ Collaboration \cite{Expt}.  One key point is
that this method for extracting $\gamma$ in fact applies to each point
in the Dalitz plot. However, the value of $\gamma$ is independent of
momentum, so that the method really represents {\it many} independent
measurements of $\gamma$. A preliminary analysis presented in
Ref.~\cite{BBCKM} considered a naive average over all such
measurements. In this letter we improve upon this and perform a
combined likelihood fit to extract $\gamma$ from multiple Dalitz-plot
points.

We begin by briefly reviewing the principal results of
Refs.~\cite{3body1,3body2}. There are three ingredients that permit
the extraction of weak phases from three-body charmless $B$
decays. First, the decay amplitudes can be expressed in terms of
diagrams. These are similar to those of two-body $B$ decays
\cite{GHLR}, except that here it is necessary to ``pop'' a quark pair
from the vacuum. The three-body diagrams are described in detail in
Ref.\ \cite{3body1}. (As we consider $\btos$ transitions, the $B^+$
decay amplitude can receive a contribution from the annihilation
diagram. This is neglected.) Note that, unlike the two-body diagrams,
the three-body diagrams are momentum dependent.

Second, it is possible to fix the symmetry of the final state. This is
done using the Dalitz plot of $B \to P_1 P_2 P_3$ (the $P_i$ are
pseudoscalar mesons) \cite{3body1}. Denoting by $p_i$ the momentum of
each $P_i$, one defines the three Mandelstam variables $s_{ij} \equiv
\left( p_i + p_j \right)^2$. These are not independent, but obey
$s_{12} + s_{13} + s_{23} = m_B^2 + m_1^2 + m_2^2 + m_3^2$. Now, the
Dalitz plot is given in terms of two Mandelstam variables, say
$s_{12}$ and $s_{13}$. The key point is that the experimental
Dalitz-plot analysis allows one to reconstruct the decay amplitude
${\cal M}(B \to P_1 P_2 P_3)(s_{12},s_{13})$. The amplitude for a
state with a given symmetry is then found by applying this symmetry to
${\cal M}(s_{12},s_{13})$.  This amplitude is used to compute all
(momentum-dependent) observables for the decay. For example, the final
state $\ks \pi^+ \pi^-$ has CP $+$ if the $\pi^+\pi^-$ pair is
symmetrized. The amplitude for this state is $[{\cal M}(s_{12},s_{13})
  + {\cal M}(s_{13},s_{12})]/\sqrt{2}$.

Third, in Ref.\ \cite{3body2} it was shown that, as is the case in
two-body decays \cite{NRGPY}, under flavor SU(3) there are relations
between the electroweak penguin (EWP) and tree diagrams for $\btos$
transitions. These take the simple form
\beq
P'_{EWi} = \kappa T'_i ~,~~ P^{\prime C}_{EWi} = \kappa C'_i ~~ (i=1,2) ~~;~~~~
\kappa \equiv - \frac{3}{2} \frac{|\lambda_t^{(s)}|}{|\lambda_u^{(s)}|}
\frac{c_9+c_{10}}{c_1+c_2} ~,
\eeq
where the $c_i$ are Wilson coefficients and $\lambda_p^{(s)}=V^*_{pb}
V_{ps}$.  Note: the EWP-tree relations hold only for the state that is
fully symmetric under exchanges of the final-state particles. However,
the amplitude for this state can be found as described above using the
Dalitz plot:
\bea \label{fulsym}
{\cal M}_{\rm fs} &=&
\frac{1}{\sqrt{6}} \left[ {\cal M}(s_{12},s_{13}) + {\cal M}(s_{13},s_{12})
+ {\cal M}(s_{12},s_{23}) \right. \nn\\
&& \hskip2truecm \left.+~{\cal M}(s_{23},s_{12}) + {\cal M}(s_{23},s_{13})
+ {\cal M}(s_{13},s_{23}) \right]~,~~
\eea
where the subscript ``fs'' stands for ``fully symmetric.''

We now describe the method proposed in Ref.~\cite{3body3} for
extracting the weak phase $\gamma$ from $\btokpipi$ and $\btokkk$
decays . The following five decays are considered: $\bd \to
K^+\pi^0\pi^-$, $\bd \to K^0\pi^+\pi^-$, $B^+ \to K^+\pi^+\pi^-$, $\bd
\to K^+ K^0 K^-$, and $\bd \to K^0 K^0 \kbar$. In writing the
amplitudes for these five processes in terms of diagrams, we note the
following. For $\btokpipi$ decays, the quark pair popped from the
vacuum is $u{\bar u}$ or $d{\bar d}$ (under isospin, these diagrams
are equal), while the $\btokkk$ decays may have a popped $s{\bar s}$
pair.  Now, the imposition of the EWP-tree relations assumes flavor
SU(3) symmetry.  But this also implies that diagrams with a popped
$s{\bar s}$ quark pair are equal to those with a popped $u{\bar u}$ or
$d{\bar d}$.  In other words, under flavor-SU(3) symmetry the diagrams
in $\btokkk$ decays are the same as those in $\btokpipi$ decays.

Note, however, that flavor-SU(3) symmetry is not exact. It is
therefore important to keep track of a possible difference between
$\btokpipi$ and $\btokkk$ decays.

The amplitudes for the five decays in terms of diagrams are given in
Ref.~\cite{3body3}.  We define the following four effective
diagrams:
\bea
a \equiv - {\tilde P}'_{tc} + \kappa \left(\frac23 T'_1 + \frac13 C'_1
+ \frac13 C'_2 \right) ~,~~~ \nn\\
b \equiv T'_1 + C'_2 ~,~~
c \equiv T'_2 + C'_1 ~,~~
d \equiv T'_1 + C'_1 ~.
\label{effdiag}
\eea
The decay amplitudes can now be written in terms of five diagrams, $a$-$d$
and ${\tilde P}'_{uc}$:
\bea
\label{effamps}
2 A(\bd \to K^+\pi^0\pi^-)_{\rm fs} &=& b e^{i\gamma} - \kappa c ~, \nn\\
\sqrt{2} A(\bd \to K^0\pi^+\pi^-)_{\rm fs} &=& -d e^{i\gamma} - {\tilde P}'_{uc} e^{i\gamma} - a + \kappa d ~, \nn\\
\sqrt{2} A(B^+ \to K^+ \pi^+ \pi^-)_{\rm fs} &=& -c e^{i\gamma} -{\tilde P}'_{uc} e^{i\gamma} - a + \kappa b ~, \nn\\
\sqrt{2} A(\bd \to K^+ K^0 K^-)_{\rm fs} &=& \alpha_{SU(3)} (-c e^{i\gamma} -{\tilde P}'_{uc} e^{i\gamma} - a + \kappa b ) ~, \nn\\
A(\bd \to K^0 K^0 \kbar)_{\rm fs} &=& \alpha_{SU(3)} ({\tilde P}'_{uc} e^{i\gamma} + a ) ~,
\eea
where $\alpha_{SU(3)}$ measures the amount of flavor-SU(3) breaking.

In the flavor-SU(3) limit ($|\alpha_{SU(3)}| = 1$), $A(B^+ \to K^+
\pi^+ \pi^-)_{\rm fs} = A(\bd \to K^+ K^0 K^-)_{\rm fs}$, so that the
$B^+$ decay does not furnish any new information. The remaining four
amplitudes depend on ten theoretical parameters: five magnitudes of
diagrams, four relative strong phases, and $\gamma$. But in principle
experiment can measure eleven observables: the decay rates and direct
CP asymmetries for the four $\bd$ decays, and the indirect CP
asymmetries of $\bd \to K^0\pi^+\pi^-$, $\bd \to K^+ K^0 K^-$ and $\bd
\to K^0 K^0 \kbar$. With more observables than theoretical parameters,
$\gamma$ can be extracted from a fit. Furthermore, if one allows for
SU(3) breaking ($|\alpha_{SU(3)}| \ne 1$), we can add two more
observables: the decay rate and direct CP asymmetry for the $B^+$
decay. In this case it is possible to extract $\gamma$ even with the
inclusion of $|\alpha_{SU(3)}|$ as a fit parameter.

As has been stressed above, the diagrams and observables are both
momentum dependent.  Thus, the extraction of $\gamma$ can be performed
at each point in the Dalitz plot. However, the value of $\gamma$ is
independent of momentum, so that we really have {\it many} independent
measurements of $\gamma$ (up to experimental correlations between
different parts of the Dalitz plot). When these are appropriately
combined, the statistical error can be reduced.

We are now in a position to apply this method for extracting $\gamma$
to real experimental data.  \babar\ has measured the Dalitz plots of
the five $\btokpipi$ and $\btokkk$ decays \cite{Expt}. The first step
in performing a fit is to collect the observables. This is done as
follows. An isobar model is used to analyze the three-body Dalitz
plots. Here the decay amplitude is expressed as the sum of a
non-resonant and several intermediate resonant contributions:
\beq
{\cal M} (s_{12}, s_{13}) = {\cal N}_{\rm DP}\sum\limits_j c_j e^{i \theta_j} F_j
(s_{12}, s_{13})~,
\eeq
where the index $j$ runs over all contributions. Each contribution is
expressed in terms of isobar coefficients $c_j$ (magnitude) and
$\theta_j$ (phase), and a dynamical wave function $F_j$. ${\cal
  N}_{\rm DP}$ is a normalization constant. The $F_j$ take different
forms depending on the contribution. The $c_j$ and $\theta_j$ are
extracted from a fit to the Dalitz-plot event distribution.  With the
amplitude in hand, the observables can be constructed at each point in
the Dalitz plot, and a fit can then be performed.

Such isobar analyses were performed by \babar\ for each of the five
three-body decays of interest \cite{Expt}. The isobar coefficients
found, together with their assumed wave functions ($F_j$), allow us to
reconstruct the amplitude for each three-body decay as a function of
the relevant Mandelstam variables. We have chosen the normalization
constant such that the integral of $|{\cal M}|^2$ over the
kinematically-allowed Dalitz-plot phase space gives the experimental
branching fraction (${\cal B}_{\rm Exp}$).  We then construct ${\cal
  M}_{\rm fs}$ using Eq.~(\ref{fulsym}). This process is repeated with
the information available for the CP-conjugate process, where we
construct $\overline{{\cal M}}_{\rm fs}$. The experimental observables
are then obtained as follows:
\bea \label{XYZdef}
X(s_{12}, s_{13}) &=& |{\cal M}_{\rm fs}(s_{12}, s_{13})|^2 + |\overline
{{\cal M}}_{\rm fs}(s_{12}, s_{13})|^2~, \nn \\
Y(s_{12}, s_{13}) &=& |{\cal M}_{\rm fs}(s_{12}, s_{13})|^2 - |\overline
{{\cal M}}_{\rm fs}(s_{12}, s_{13})|^2~, \nn \\
Z(s_{12}, s_{13}) &=& {\rm Im}\left[{\cal M}^*_{\rm fs}(s_{12}, s_{13})
~\overline{{\cal M}}_{\rm fs}(s_{12}, s_{13})\right]~.
\eea
Here, $X$, $Y$ and $Z$ are, respectively, the effective CP-averaged
branching ratio, the direct CP asymmetry, and the indirect CP
asymmetry. These may be constructed for every point on any Dalitz
plot.  However, when the final state has a specific flavor, such as in
the case of $B^0\to K^+\pi^0\pi^-$, the quantity $Z$ has no physical
meaning and is therefore left out of our analysis.

In order to obtain the experimental errors on these quantities, we
vary the input isobar coefficients over their $1\sigma$
statistical-uncertainty ranges.  We include the correlations between
these coefficients when they are given in the papers of
Ref.~\cite{Expt}.

\begin{figure}[!htbp]
    \centering
         \includegraphics[scale=0.5]{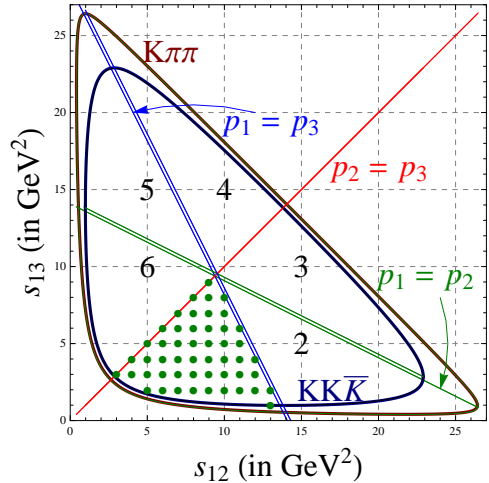}
\caption{Kinematic boundaries and symmetry axes of $\btokpipi$ and
$\btokkk$ Dalitz plots. The symmetry axes divide each plot into six
zones, five of which are marked 2-6. The fifty dots in the region
of overlap of the first of six zones from all Dalitz plots are used
for the $\gamma$ measurement.
\label{DPbdry}}
\end{figure}

In addition, note that, since the amplitudes used to construct these
observables are fully symmetric under the interchange of the three
Mandelstam variables, for any given point on a Dalitz plot there will
be five other points where the extracted $X$, $Y$ and $Z$ take
identical values, and hence do not provide any new information.  In
order to avoid counting the same information multiple times, we
therefore divide each Dalitz plot into six zones by its three axes of
symmetry, and use information only from one zone. This is shown in
Fig.~\ref{DPbdry}, where we select the dotted zone for our
calculations.

The next step is to pick the points on the Dalitz plot where the
observables can be evaluated. The idea is to choose the maximum number
of points for which the observables evaluated at these points are
independent of one another.  Ideally, with enough data (and a perfect
apparatus), every point in the region of overlap can be treated as an
independent source for measuring $\gamma$. In practice, however, the 
maximum number of independent points is limited by the number of events 
observed in the three-body decays. Here we pick a grid with an equal 
spacing of 1 GeV$^2$ between two consecutive points (this spacing is 
chosen arbitrarily). We find that there are fifty such points. In the 
experimental data, the process with the smallest statistics is $\bd 
\to K^0 K^0 \kbar$, for which \babar\ has reported $200\pm15$ events 
\cite{Expt}. Our choice of spacing is consistent with this number of 
events.  We perform a maximum likelihood fit to the observables at 
these fifty points to obtain $\gamma$. Note that this is just an example.  
Since the final value of $\gamma$ is essentially the average over all 
points, its error scales simply as $1/\sqrt{N}$.  The maximum value 
that $N$ can take is limited by the available experimental statistics.

With direct access to the data, a more accurate analysis for
determining $N$ is possible. The data can be separated into bins in
each of the two Mandelstam variables; an optimal bin size, not
necessarily uniform over the Dalitz plot, can be suitably chosen. Note
that it is necessary to choose identical binning for all the processes
involved.  Observables in each of the $N$ bins can then be used as an
independent source of measuring $\gamma$.

Although it is possible in principle to measure both the direct and
indirect CP asymmetries in $B^0 \to \ks\ks\ks$, their measurement is
currently statistics limited. The experimental Dalitz-plot analysis
done by \babar\ makes no distinction between the amplitude and its CP
conjugate. That is, they take $A(\bd\to\ks\ks\ks) =
A(\bdbar\to\ks\ks\ks)$. This has two consequences.  First, $Y$ and $Z$
vanish for every point of the Dalitz plot.  Second, this requires that
${\tilde P}'_{uc}$ be set to zero in Eq.~(\ref{effamps}).  The removal
of an equal number of unknown parameters (amplitude and phase of
${\tilde P}'_{uc}$) and observables does not affect the viability of
the method described above.

With the observables in hand, we now perform a maximum likelihood
analysis for extracting $\gamma$.  For each of the fifty points in
region 1 of Fig.~\ref{DPbdry}, we construct the $-2 \Delta \ln {\rm
  L}(\gamma)$ function, where ${\rm L}$ represents the likelihood,
which we then minimize over all the hadronic parameters for that
point.  Since we have assumed the observables of Eq.~(\ref{XYZdef}) to
be uncorrelated, the fifty points are independent, so that the sum of
log-likelihood functions over all points gives us a joint likelihood
distribution. The local minima of this function are then identified as
the most-likely central values of $\gamma$.  The values of $\gamma$
for which there is a unit shift along the vertical axis of the $-2
\Delta \ln {\rm L}(\gamma)$ vs $\gamma$ plot represent the $1\sigma$
range corresponding to each central value.

\begin{figure*}[!htbp]
    \centering
\includegraphics[scale=1.1]{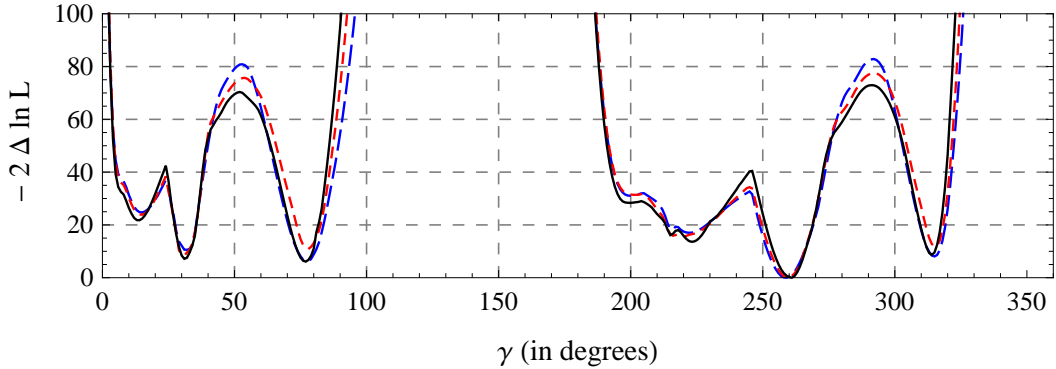}
\caption{Results of maximum-likelihood fits. The solid (black) curve
represents the fit assuming flavor-SU(3) symmetry. The short dashes (red)
represent the fit where flavor-SU(3) breaking is fixed by a point-by-point
comparison of Dalitz plots for $B^+\to K^+\pi^+\pi^-$ and $B^0\to K^+ K^0
K^-$. The long dashes (blue) represent the fit with inputs from five Dalitz
plots and an extra hadronic fit parameter $|\alpha_{SU(3)}|$. \label{maxlik}}
\end{figure*}

We perform the likelihood maximization fit in three different ways and
plot our results in Fig.\ \ref{maxlik}. We first consider the scenario
in which flavor SU(3) is a good symmetry. That is, we fix $|\alpha_{SU(3)}|
= 1$; our analysis involves only the four $B^0$ decay channels. The
most-likely values of $\gamma$ obtained in this way are listed under Fit 1
in Table \ref{gamma}.

Second, we allow for SU(3) breaking and treat it as follows. We compare
the Dalitz plots for the two processes $B^+\to K^+\pi^+\pi^-$ and $B^0
\to K^+ K^0 K^-$ point by point. Theoretically, the amplitudes for these
processes differ only by the parameter $\alpha_{SU(3)}$. The ratio of $X$'s
constructed from the two Dalitz plots then gives us $|\alpha_{SU(3)}|^2$.
(Note that a similar ratio constructed from the $Y$'s has an enormous
error due to the smallness of $Y$. We are therefore unable to extract any
interesting physical information from such a ratio.) Averaged over the
fifty points we find $|\alpha_{SU(3)}| = 0.97 \pm 0.05$. This shows that,
on average, SU(3) breaking is small. We use $|\alpha_{SU(3)}|$ found in
this way to correct the observables from the $\btokkk$ Dalitz plots and
use the corrected numbers in a new maximum-likelihood analysis for finding
$\gamma$. We present the results under Fit 2 in Table \ref{gamma}.

In the third maximum-likelihood analysis, we consider observables from
all five Dalitz plots but now include $|\alpha_{SU(3)}|$ as an additional
unknown hadronic parameter. The results from this method are listed
under Fit 3 in Table \ref{gamma}.

\begin{table}[!htbp]
\renewcommand\arraystretch{1.2}
\setlength{\abovecaptionskip}{5pt}         
\setlength{\belowcaptionskip}{-5pt}        
\caption{Most likely values of $\gamma$ (in degrees) extracted from Fig.\
\ref{maxlik}. Results are obtained using the three different fitting methods
as explained in the text. \label{gamma}}
\begin{center}
\begin{tabular}{c c c c c} \hline \hline
Solution & ~~Fit 1~~ & ~~Fit 2~~ & ~~Fit 3~~ \\ \hline \hline
 I   &$31^{+2}_{-1}$ &$31^{+1}_{-2}$ & $32\pm2$ \\
 II  &$77\pm2$       &$78 \pm2$      & $77\pm2$ \\
 III &$261^{+2}_{-3}$&$259^{+3}_{-2}$& $259^{+2}_{-3}$ \\
 IV  &$314\pm2$      &$315\pm2$      & $315\pm2$ \\
\hline
\end{tabular}
\end{center}
\end{table}
\setlength{\abovecaptionskip}{10pt}  
\setlength{\belowcaptionskip}{0pt}   

The maximum-likelihood analysis indicates that, in each of the three
methods described above, the data favor four distinct
discretely-ambiguous values of $\gamma$. (Due to the fact that the fit
involves nonlinear equations, it is not surprising to find multiple
solutions for $\gamma$.)  In Table \ref{gamma} we present the
most-likely values of $\gamma$ extracted using these three methods. It
is evident from the results that the inclusion of an SU(3)-breaking
parameter $|\alpha_{SU(3)}|$ shifts the preferred values of $\gamma$
by only a tiny amount. This indicates that the leading-order effects
of flavor-SU(3) breaking are well under control in three-body $B$
decays. While one cannot completely remove this source of theoretical
error from our analysis, the uncertainties are rather small.

Even though there are four preferred values of $\gamma$, in all cases
the error is small, 2-3$^\circ$. Although this may seem surprising at
first sight, it really is not when one remembers that there are, in
fact, fifty independent measurements of $\gamma$. Roughly speaking, if
each measurement has an error of $\pm 20^\circ$ \cite{BBCKM}, which is
somewhat larger than other methods, then when we take a naive average,
we divide the error by $\sqrt{50}$, giving a final error of $\sim
3^\circ$. And, as noted above, if the number of independent points in
the Dalitz plot is not fifty, but twenty (for example), the error will
be increased by about $\sqrt{50}/\sqrt{20}$.

Because the observables at different points are all computed using the
same isobar coefficients, there is a certain level of correlation,
reducing the degree to which these points are independent. That is,
the effective value of $N$ is decreased, leading to an increased error
on $\gamma$. We refer to this as the ``correlation error.'' A precise
estimate of the correlation error requires detailed information about 
the statistical and systematic errors on the isobar parameters, as well 
as the correlations between them. Even if such information were completely 
available, a full analysis would involve a multi-parameter fit requiring 
computational power that is well beyond the scope of our present analysis.

There are other sources of error that have not been included in our
(simple) analysis. First, and most importantly, all errors considered
to this point have been entirely statistical -- the systematic error
has not been included. The reason is that only statistical errors were
given for the isobar coefficients in the \babar\ papers of
Ref.~\cite{Expt}.  Second, we have only taken leading-order
flavor-SU(3) breaking into account.  Higher-order flavor-SU(3)
breaking may arise due to the nonzero mass difference between pions
and kaons, and between intermediate resonances. This said, the error
due to leading-order SU(3) breaking is evidently small. It is unlikely
that the error due to higher-order SU(3) breaking is larger.

To summarize: we have demonstrated that it is possible to cleanly
extract $\gamma$ from $\btokpipi$ and $\btokkk$ decays, and we find
four most-likely values. Three of these -- $32^\circ$, $259^\circ$,
and $315^\circ$ -- are in disagreement with the SM (is this a ``$K \pi
\pi$-$KK{\bar K}$ puzzle''?). However, one solution -- $77^\circ$ --
is consistent with the SM. In all cases, although we find a small
error, we have made a number of assumptions about the data in
performing the analysis, and several sources of error have been
ignored. The full error on $\gamma$ must be determined in order to
judge the efficacy of this method. This can only be done with direct
access to the data, and hopefully this procedure for extracting
weak-phase information from three-body $B$ decays will be incorporated
into the programs of future experiments (e.g., a super $B$ factory,
perhaps LHCb).

\bigskip
\noindent
{\bf Acknowledgments}: A special thank you goes to E. Ben-Haim for his
important input to this project. We also thank J. Charles, M. Gronau,
N. Rey-Le Lorier, J. Rosner and J. Smith for helpful
communications. BB would like to thank G. Bell and WG IV of CKM
2012. This work was facilitated in part by the workshop {\it New
  Physics from Heavy Quarks in Hadron Colliders,} which was sponsored
by the University of Washington and supported by the DOE under
contract DE-FG02-96ER40956. This work was financially supported by
NSERC of Canada (BB, DL) and by FQRNT du Qu\'ebec (MI).

\end{document}